\documentclass{article}
\usepackage{graphicx}
\usepackage{siunitx}
\usepackage{amsbsy}
\usepackage{amsmath,bm}
\usepackage{amssymb}
\usepackage{amsfonts}
\usepackage{mathtools}
\usepackage[bottom]{footmisc}
\usepackage{cite}
\usepackage{environ}
\usepackage{subcaption}

\usepackage{tikz}
\usepackage{pgfplots}
\usepackage{xcolor}
\usepackage{tudacolors}

\usepackage{tikzscale} 
\graphicspath{{tikz/}}

\usepackage{url}

\begin{document}
\title{Mortar Thin Shell Approximation for Analysis of Superconducting Accelerator Magnets}
\author{Robert Hahn$^{1}$
	\and
	Erik Schnaubelt$^{1,2}$
	\and
	Mariusz Wozniak$^{2}$
	\and
	Christophe Geuzaine$^{3}$
	\and
	Sebastian Schöps$^{1}$
}
\date{\small
	$^1$Technical University of Darmstadt, Darmstadt, Germany \\
	\texttt{robert.hahn@tu-darmstadt.de, sebastian.schoeps@tu-darmstadt.de}\\[2ex]%
	$^2$CERN, Meyrin, Switzerland \\
	\texttt{erik.schnaubelt@cern.ch, mariusz.wozniak@cern.ch}\\[2ex]%
	$^3$University of Liège, Liège, Belgium \\
	\texttt{cgeuzaine@uliege.be}
}

\maketitle              

\begin{abstract}
	Thin layers can lead to unfavorable meshes in a finite element (FE) analysis. Thin shell approximations (TSAs) avoid this issue by removing the need for a mesh of the thin layer while approximating the physics across the layer by an interface condition. Typically, a TSA requires the mesh of both sides of the TSA interface to be conforming. To alleviate this requirement, we propose to combine mortar methods and TSAs for solving the heat equation. The mortar TSA method's formulation is derived and enables an independent discretization of the subdomains on the two sides of the TSA depending on their accuracy requirements. The method is verified by comparison with a reference FE solution of a thermal model problem of a simplified superconducting accelerator magnet.
	\\\textbf{keywords: finite element method, thin shell approximation, mortar method}
\end{abstract}
\section{Introduction}%
\label{hahn:sec_intro}
Superconducting electromagnets are used for example in particle accelerators.
To ensure operational safety, multiphysical --- in particular thermal and electromagnetic --- models of these magnets are simulated.
Unfortunately, the magnets often include thin volumetric layers, such as electrical insulation or turn-to-turn contacts. The existence of these layers presents a challenge for  numerical simulations, as e.g.\ a naive finite element (FE) discretization can lead to unfavorable meshes, either due to a high number of degrees of freedom or low-quality mesh elements~\cite{Driesen_2001aa}.
To alleviate this problem, thin shell approximations~(TSAs) collapse the thin volumetric layer into a surface and approximate the physics inside the thin layer, commonly leading to field discontinuities across the surface.
\par
For the model computational domain $\Omega$, we assume a thin volumetric layer $\Omega_\mathrm{i}$ called the \emph{internal} subdomain,
separated from the two \emph{external} subdomains $\Omega_{\mathrm{e}, 1}$ and $\Omega_{\mathrm{e}, 2}$ by the interfaces $\Gamma_1$ and $\Gamma_2$. An illustration of the configuration is shown in Fig.~\ref{hahn:fig_compDomain}.
Typically, TSAs link the problem formulations on $\Omega_{\mathrm{e}, 1}, \Omega_{\mathrm{e}, 2}$ and $\Omega_\mathrm{i}$ by enforcing equality of the respective interface terms in a strong sense. This procedure requires conforming meshes for $\Gamma_1$ and $\Gamma_2$~\cite{Driesen_2001aa,Alves_2021aa,Schnaubelt_2023aa}. However, using different discretization levels of $\Omega_{\mathrm{e}, 1}$ and $\Omega_{\mathrm{e}, 2}$ (as shown on the right in Fig.~\ref{hahn:fig_compDomain}) can be advantageous in the engineering context.
Mortar methods enable the use of different discretizations on different subdomains. Equality on the interfaces is then imposed not in a strong sense, but by introducing Lagrange multipliers~\cite[Section 2.5.1]{Quarteroni_1999aa}.
The proposed \emph{mortar TSA} method enables the use of TSAs with non-conforming meshes by combining TSAs with the mortar method.
\begin{figure}[t]
	\centering
	\includegraphics[width=0.75 \columnwidth]{top_view_rect}
	\caption{Cross-section of computational domain $\Omega$ for the meshed reference (left) and mortar TSA (right), which also shows a sketch of a non-conforming mesh. The reference mesh is omitted for the sake of a clear visualization.
	}%
    \label{hahn:fig_compDomain}
\end{figure}

\section{Classical Weak Formulation}

The starting point for introducing the mortar TSA is a weak formulation with conforming mesh. We solve the heat equation: \\Find $T \in V_g = \left\{u \in H^1 \! \left(\Omega\right) : u = g \text{ on } \Gamma_\mathrm{Dir} \subset \partial\!\Omega \right\}$ s.t.
\begin{equation}
	\begin{split}
		&\left(\kappa \nabla T, \nabla T'\right)_{\Omega} +  \left( C_\mathrm{V} \, \partial_t T, T' \right)_{\Omega} = \left( Q, T' \right)_{\Omega} + \big \langle\vec{n} \cdot \kappa \nabla T, T' \big \rangle_{\partial\! \Omega} \, \forall T' \in V_0\;\text{.}
	\end{split}%
	\label{hahn:eq_weak-heat}
\end{equation}
Herein, $T$ is the temperature, $\kappa$ the thermal conductivity, $C_\mathrm{V}$ the volumetric heat capacity, $Q$ a heat source, $\vec{n}$ the outward-oriented normal vector and $H^1 \! \left(\Omega\right)$ the space of square integrable ($L^2$) functions with square integrable weak gradient.
The $L^2$-inner product, that is the integral of the scalar product of two functions, is denoted by $(\cdot,\cdot)_\Omega$ for three dimensional and by $\langle \cdot, \cdot\rangle_{\partial\!\Omega}$ for two dimensional regions.
Suitable boundary conditions (BCs) have to be defined on $\partial\!\Omega$. In addition to the Dirichlet-BCs imposed in the space of permissible solutions $V_g$, we use adiabatic homogeneous Neumann and Robin BCs, that is
\begin{alignat}{2}
	\vec{n} \cdot \left(\kappa \nabla T\right) &= 0 \quad &\text{on  } \Gamma_\mathrm{Neu} \subset \partial\!\Omega\;\text{,}
	\label{hahn:eq_neumann} \\
	\vec{n} \cdot \left(\kappa \nabla T\right) &= h \left(T_\mathrm{ref} - T\right) \quad &\text{on  } \Gamma_\mathrm{Rob} \subset \partial\!\Omega\;\text{,}
	\label{hahn:eq_robin}
\end{alignat}
with a reference temperature $T_\mathrm{ref}$ and the heat transfer coefficient $h$.

\section{Mortar Thin Shell Formulation}%
\label{hahn:sec_formulation}
In order to introduce the mortar TSA and starting from~\eqref{hahn:eq_weak-heat}, the external subdomain $\Omega_\mathrm{e} \coloneqq \Omega_{\mathrm{e}, 1} \cup \, \Omega_{\mathrm{e}, 2}$ and the internal subdomain $\Omega_{\mathrm{i}}$ are considered separately.
They are connected via the interface contributions
\begin{equation}
	\big \langle \vec{n}_1 \cdot \kappa \nabla T, T'\big \rangle_{\Gamma_1} + \big \langle \vec{n}_2 \cdot \kappa \nabla T, T'\big \rangle_{\Gamma_2} \eqqcolon B(T, T')\;\text{,}
	\label{hahn:eq_interface_term}
\end{equation}
where the normal vectors $\vec{n}_1, \vec{n}_2$ are oriented outwards w.r.t. $\Omega_\mathrm{i}$.
Preserving the possibility to enforce Robin-BCs on parts of $\partial\!\Omega_{\mathrm{e}}\setminus\{\Gamma_1 \cup \Gamma_2\}$ and assuming either Dirichlet or homogeneous Neumann BCs everywhere else, we find the two sub-problems
\begin{align}
	\begin{split}
	\left(\kappa \nabla T, \nabla T'\right)_{\Omega_{\mathrm{e}}}
	&+ \left( C_\mathrm{V} \, \partial_t T, T' \right)_{\Omega_{\mathrm{e}}}
	+ \big \langle h(T-T_\mathrm{ref}), T' \big \rangle_{\Gamma_\mathrm{Rob}}\\
	\quad =\left( Q, T' \right)_{\Omega_{\mathrm{e}}}
	&- B(T, T')\;\text{,}%
	\end{split}
	\label{hahn:eq_external_problem}\\
	\left(\kappa \nabla T, \nabla T'\right)_{\Omega_{\mathrm{i}}}
	&+ \left( C_\mathrm{V} \, \partial_t T, T' \right)_{\Omega_{\mathrm{i}}}
	=\left( Q, T' \right)_{\Omega_{\mathrm{i}}} + B(T, T')\;\text{,}%
	\label{hahn:eq_internal_problem}
\end{align}
where the \emph{external problem}~\eqref{hahn:eq_external_problem} is defined on $\Omega_{\mathrm{e}}$, and the \emph{internal problem}~\eqref{hahn:eq_internal_problem} is defined on $\Omega_{\mathrm{i}}$.
For the TSA, we then seek an approximate solution $\hat{T}$ of the internal problem~\eqref{hahn:eq_internal_problem}, for which we consider the heat equation on $\hat{\Omega_\mathrm{i}}$, a (potentially approximate) representation of the original thin volumetric layer $\Omega_\mathrm{i}$.
We subdivide the domain $\hat{\Omega_\mathrm{i}}$ into $N$ disjoint layers~\cite{Alves_2021aa}, that is
$\hat{\Omega}_\mathrm{i} = \bigcup_{k=1}^N \hat{\Omega}_\mathrm{i}^{(k)}$
and introduce a local coordinate system $\vec{u}, \vec{v}, \vec{w}$ with $\vec{w}$ normal to the virtual interface $\hat{\Gamma}$ and $\vec{u}, \vec{v}$ in tangential direction.
We then use a tensor product ansatz for the layers, i.e.
\begin{equation}
	\hat{\Omega}_\mathrm{i}^{(k)} = \hat{\Gamma} \times [w_{k-1}, w_k]\;\text{,}
	\label{hahn:eq_tensor-product-ansatz}
\end{equation}
as is illustrated in Fig.~\ref{hahn:fig_virtual_disc}.
\begin{figure}[t]
	\centering
	\includegraphics[width=0.8\columnwidth]{virtual_disc}
	\caption{Tensor product discretization of $\hat{\Omega}_\mathrm{i}$.}%
	\label{hahn:fig_virtual_disc}
\end{figure}
Denoting by $\hat{T}^{(k)} \coloneqq \hat{T}|_{\hat{\Omega_\mathrm{i}}^{(k)}}$ the solution within the layer $\smash{\hat{\Omega}_\mathrm{i}^{(k)}}$ and by
$\smash{\hat{\Gamma}_0} = \smash{\hat{\Gamma}} \times w_0$ and $\smash{\hat{\Gamma}_N} = \smash{\hat{\Gamma}} \times w_N$
the outer boundary surfaces of $\smash{\hat{\Omega}}_\mathrm{i}$, the (approximate) internal problem can be rewritten as
\begin{align}
		&\hat{B}(\hat{T}, \hat{T}')
		= \big \langle \vec{n}_1 \cdot \kappa \nabla \hat{T}^{(0)} , \hat{T}'^{(0)}\big \rangle_{\hat{\Gamma}_0}
		+ \big \langle \vec{n}_2 \cdot \kappa \nabla \hat{T}^{(N)} , \hat{T}'^{(N)}\big \rangle_{\hat{\Gamma}_N} \label{hahn:eq_interface_term_hat} \\
		&\quad= \sum_{k=1}^N
		\bigg\{
		\Bigl(\kappa \nabla \hat T^{(k)}, \nabla \hat T'^{(k)}\Bigr)_{\hat\Omega_{\mathrm{i}}^{(k)}}  
		+ \Bigl( C_\mathrm{V} \, \partial_t \hat T^{(k)}, \hat T'^{(k)} \Bigr)_{\hat\Omega_{\mathrm{i}}^{(k)}}
		- \Bigl( Q, \hat T'^{(k)} \Bigr)_{\hat\Omega_{\mathrm{i}}^{(k)}}
		\bigg\}
		\;\text{.} \nonumber
\end{align}
This approximates~\eqref{hahn:eq_interface_term}.
We then assume a product decomposition within each layer, i.e.
\begin{equation}
	\hat{T}^{(k)}(u,v,w,t) = \sum_{j = k-1}^k \hat{T}_j (u,v,t)\Psi_j (w)\;\text{,}
	\label{hahn:eq_product-decomposition}
\end{equation}
with the degrees of freedom (DoFs) $\hat{T}_j$ of the TSA supported on $\hat{\Gamma}$ and the one dimensional basis functions $\Psi_j$ e.g.\ chosen as first-order Lagrange basis functions. As the DoFs  $\hat{T}_j$ are independent of $w$, we then use the tensor product ansatz~\eqref{hahn:eq_tensor-product-ansatz} to decompose the integration domain
\begin{equation}
	\int_{\hat{\Omega}^{(k)}} \mathrm{d}\hat\Omega = \int_{\hat{\Gamma}} \int_{w_{k-1}}^{w_k} \mathrm{d}w \,\mathrm{d}\hat\Gamma\;\text{,}
	\label{hahn:eq_integral-split}
\end{equation}
and explicitly evaluate the integrals w.r.t. $w$. Using the notations
\begin{equation}
	\begin{aligned}
		\Bigl(\widehat{\mathbf{K}}_{\kappa}^{(k)} \Bigr)_{ij} &= \widehat{K}_{\kappa, ij}^{(k)}
		&&\coloneqq \int_{w_{k-1}}^{w_k} \kappa \partial_w \Psi_j(w) \partial_w \Psi_i(w) \mathrm{d}w\;\text{,}\\
		\Bigl(\widehat{\mathbf{M}}_{\kappa}^{(k)} \Bigr)_{ij} &= \widehat{M}_{\kappa, ij}^{(k)}
		&&\coloneqq \int_{w_{k-1}}^{w_k} \kappa \Psi_j(w) \Psi_i(w)\mathrm{d}w\;\text{,}\\
		\Bigl(\widehat{\mathbf{M}}_{c_V}^{(k)} \Bigr)_{ij} &= \widehat{M}_{c_V, ij}^{(k)}
		&&\coloneqq \int_{w_{k-1}}^{w_k} c_V \Psi_j(w) \Psi_i(w) \mathrm{d}w\;\text{,}\\
		\Bigl(\hat{\mathbf{q}}^{(k)} \Bigr)_{i} &= \hat{q}_{i}^{(k)}
		&&\coloneqq \int_{w_{k-1}}^{w_k} Q \Psi_i(w) \mathrm{d}w\;\text{,}\\
	\end{aligned}
	\label{hahn:eq_1D-matrices}
\end{equation}
for the matrices corresponding to this one dimensional discretization,
we find for the terms in the (approximate) internal problem~\eqref{hahn:eq_interface_term_hat}:
\begin{align}
		\left(\kappa \nabla \hat T^{(k)}, \nabla \hat T'^{(k)}\right)_{\hat\Omega_{\mathrm{i}}^{(k)}}
		&= \sum_{j=k-1}^k \sum_{i=k-1}^k
		\Bigl \langle \widehat{K}_{\kappa, ij}^{(k)}\,\hat{T}_j,  \hat{T}_i' \Bigr \rangle_{\hat{\Gamma}}
		+ \Bigl \langle \widehat{M}_{\kappa, ij}^{(k)}\,\nabla \hat{T}_j,  \nabla \hat{T}_i' \Bigr \rangle_{\hat{\Gamma}}\;\text{,} \nonumber \\
		\left( C_\mathrm{V}\,\partial_t \hat T^{(k)}, \hat T'^{(k)} \right)_{\hat\Omega_{\mathrm{i}}^{(k)}}
		&= \sum_{j=k-1}^k \sum_{i=k-1}^k
		\Bigl \langle \widehat{M}_{C_V, ij}^{(k)} \,\partial_t \hat{T}_j, T_i' \Bigr \rangle_{\hat{\Gamma}}\;\text{,} \label{hahn:eq_layer-terms-matrix} \\
		\left( Q, \hat T'^{(k)} \right)_{\hat\Omega_{\mathrm{i}}^{(k)}}
		&= \sum_{i=k-1}^k
		\Bigl \langle \hat{q}_{i}^{(k)}, \hat{T}_i' \Bigr \rangle_{\hat{\Gamma}} \nonumber\;\text{.}
\end{align}
This yields the internal problem formulation:
\begin{align}
	&\sum_{k=1}^N \sum_{j = k-1}^k \sum_{i = k-1}^k
	\bigg\{
	\Bigl \langle  \widehat{K}_{\kappa, ij}^{(k)}\,\hat{T}_j,  \hat{T}_i' \Bigr \rangle_{\hat{\Gamma}}
	+ \Bigl \langle \widehat{M}_{\kappa, ij}^{(k)}\,\nabla \hat{T}_j,  \nabla \hat{T}_i' \Bigr \rangle_{\hat{\Gamma}}
	+ \Bigl \langle \widehat{M}_{c_V, ij}^{(k)}\, \partial_t \hat{T}_j, T_i' \Bigr \rangle_{\hat{\Gamma}} \bigg\} \nonumber \\
	&=\hat B(\hat{T}, \hat{T}')
	+ \sum_{k=1}^N \sum_{i = k-1}^k	 \Bigl \langle \hat{q}_{i}^{(k)}, \hat{T}_i' \Bigr \rangle_{\hat{\Gamma}}
	\;\text{.}
	\label{hahn:eq_internal_triple_sum}
\end{align}
The decomposition of the internal problem in integrals over $\hat{\Gamma}$ and 1D FE matrices in $[w_{k-1}, w_k]$ leads to a formulation of the internal problem in which no volumetric mesh of $\hat{\Omega}_\mathrm{i}$ is needed. A detailed derivation is found in~\cite{Schnaubelt_2023aa}.\par
The connection of the external and internal problem is established in~\cite{Schnaubelt_2023aa} by enforcing ${T|}_{\Gamma_1} = {\hat{T}|}_{\hat{\Gamma}_0}$ and ${T|}_{\Gamma_2} = {\hat{T}|}_{\hat{\Gamma}_N}$ in a strong sense. This requires \emph{conforming meshes} of $\smash{\hat{\Gamma}}$, $\Gamma_1$ and $\Gamma_2$.
We propose to circumvent this constraint by using a \emph{mortar method} to couple the external and internal problems. To this end, we introduce the two Lagrange multipliers
\begin{equation}
	\lambda_1 \in \Lambda_1 = H^{-1/2}(\Gamma_1) \text{ and }
	\lambda_2 \in \Lambda_2 = H^{-1/2}(\Gamma_2)\;\text{,}
\end{equation}
where for $j\in \{1,2\}$, $H^{-1/2}(\partial\!\Omega_{\mathrm{e},j})$ denotes the trace space of $L^2(\Omega_{\mathrm{e},j})$, and $H^{-1/2}(\Gamma_j)$ its restriction to the interface $\Gamma_j$.
To achieve weak continuity across the interfaces $\Gamma_1$ and $\Gamma_2$, we require weak equality of both temperature and heat flux for the two sides of each interface.
To enforce weak equal temperature, we introduce the additional conditions
\begin{equation}
	\begin{split}
		\big \langle T, \lambda_1' \big \rangle_{\Gamma_1}
		&= \big \langle \hat{T}, \lambda_1'\big \rangle_{\hat{\Gamma}_0}
		\quad \forall \lambda_1' \in \Lambda_1\;\text{,}\\
		\big \langle T, \lambda_2' \big \rangle_{\Gamma_2}
		&= \big \langle \hat{T}, \lambda_2'\big \rangle_{\hat{\Gamma}_N}
		\quad \forall \lambda_2' \in \Lambda_2\;\text{.}
	\end{split}
\end{equation}
Weak equality of the interface fluxes is achieved by inserting $\lambda_1$ in place of both $\vec{n}_1 \cdot \kappa \nabla T|_{\Gamma_1}$ in~\eqref{hahn:eq_interface_term} and $\vec{n}_1 \cdot \kappa \nabla \hat{T}^{(0)}|_{\hat{\Gamma}_0}$ in~\eqref{hahn:eq_interface_term_hat}.
With an analogous approach for $\lambda_2$, we thereby find the modified interface contributions
\begin{equation}
	\begin{split}
	B(T, T') &= \big \langle \lambda_1, T' \big \rangle_{\Gamma_1}
	+ \big \langle \lambda_2, T' \big \rangle_{\Gamma_2}\;\text{,}\\
	\hat B(\hat{T}, \hat{T}') &= \big \langle \lambda_1, \hat{T}'^{(0)} \big \rangle_{\hat{\Gamma}_0}
	+ \big \langle \lambda_2, \hat{T}'^{(N)} \big \rangle_{\hat{\Gamma}_N}\;\text{.}
	\label{hahn:eq_interface_contributions_modified}
	\end{split}
\end{equation}
Let us note that by choosing $\hat{\Gamma}$ conformal to $\Gamma_1$ (or $\Gamma_2$), $\lambda_1$ (or $\lambda_2$) can be eliminated.
The final formulation is recovered by inserting the modified interface contributions~\eqref{hahn:eq_interface_contributions_modified}
in the internal~\eqref{hahn:eq_internal_triple_sum} and external~\eqref{hahn:eq_external_problem} problem formulations.\par
An extension of the mortar TSA method for magnetodynamic $\vec{H}$--$\varphi$ TSA formulations~\cite{Schnaubelt_2023ab} is straightforward but requires to work with the corresponding subsets of the edge element space $H(\operatorname*{curl},\Omega)$ and their traces.

\section{Numerical Example}%
\label{hahn:sec_numerical_example}
To demonstrate the correctness of the mortar TSA method, we consider a non-linear model problem based on an accelerator magnet geometry. In this two dimensional problem, we consider two adjacent superconducting Niobium-Titanium composite cables separated by thin Kapton insulation layers, for which the mortar TSA is used. Below the cables is a gap, which is assumed to be filled with Kapton as well, followed by a steel collar.
The cables are heated via a constant heat source $Q=\SI{1e5}{\watt\per\meter\squared}$
to model the resistive losses encountered in a quench event. For the right cable, we employ a Robin boundary condition to model cryogenic cooling of the material. An illustration of the configuration at hand is given in Fig.~\ref{hahn:fig_example}.\par
\begin{figure}[h]
	\vspace{-1.25cm}
	\centering
    \includegraphics[width=0.75 \columnwidth]{tscale} \\[0.25em]
	\vspace{0.25cm}
	\begin{subfigure}{0.49\textwidth}
		\centering
		\includegraphics[width=0.9\columnwidth]{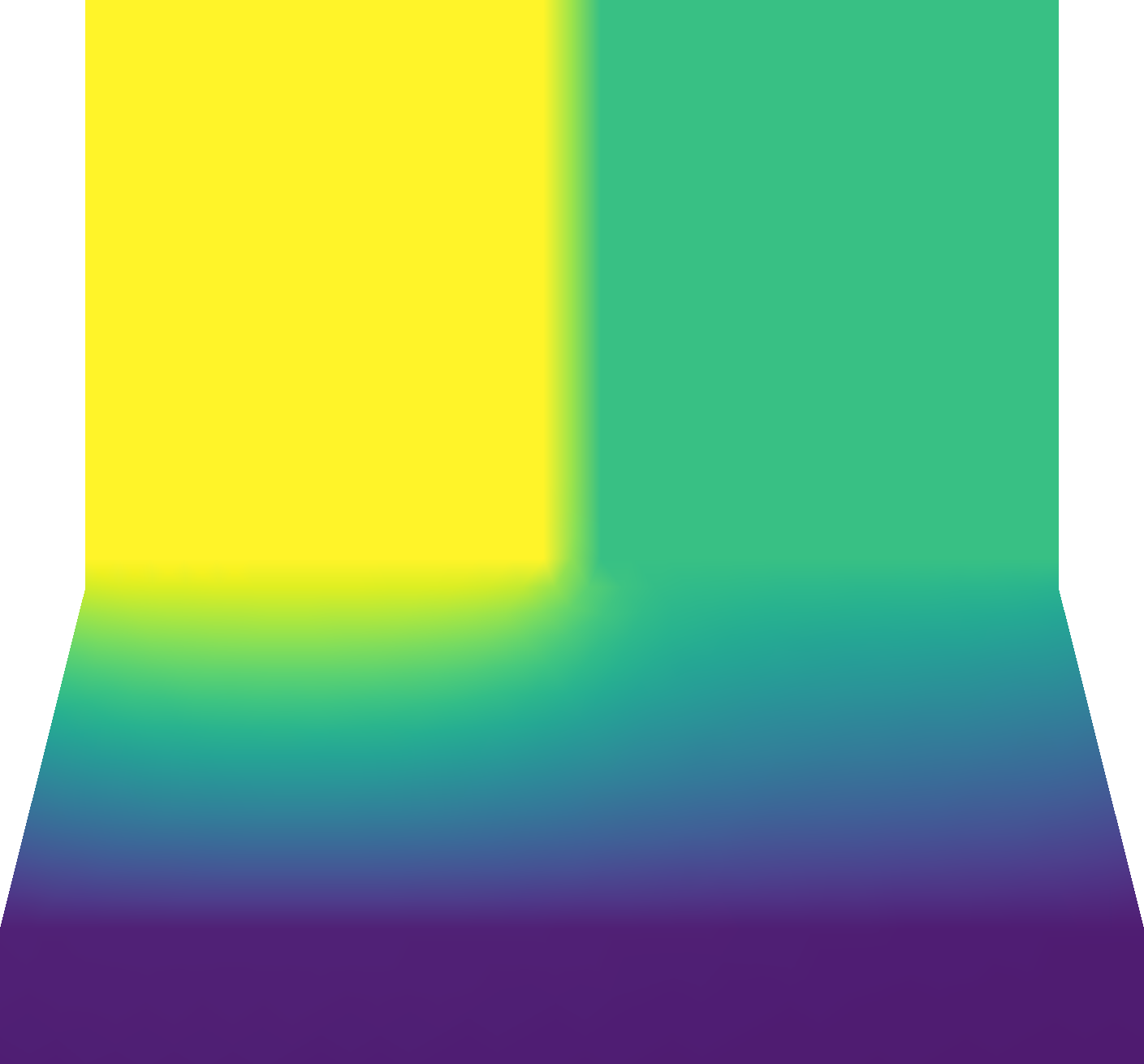}%
		\caption{Reference solution.}%
	\end{subfigure}
	\hfill
	\begin{subfigure}{0.49\textwidth}
		\centering
		\includegraphics[width=0.9\columnwidth]{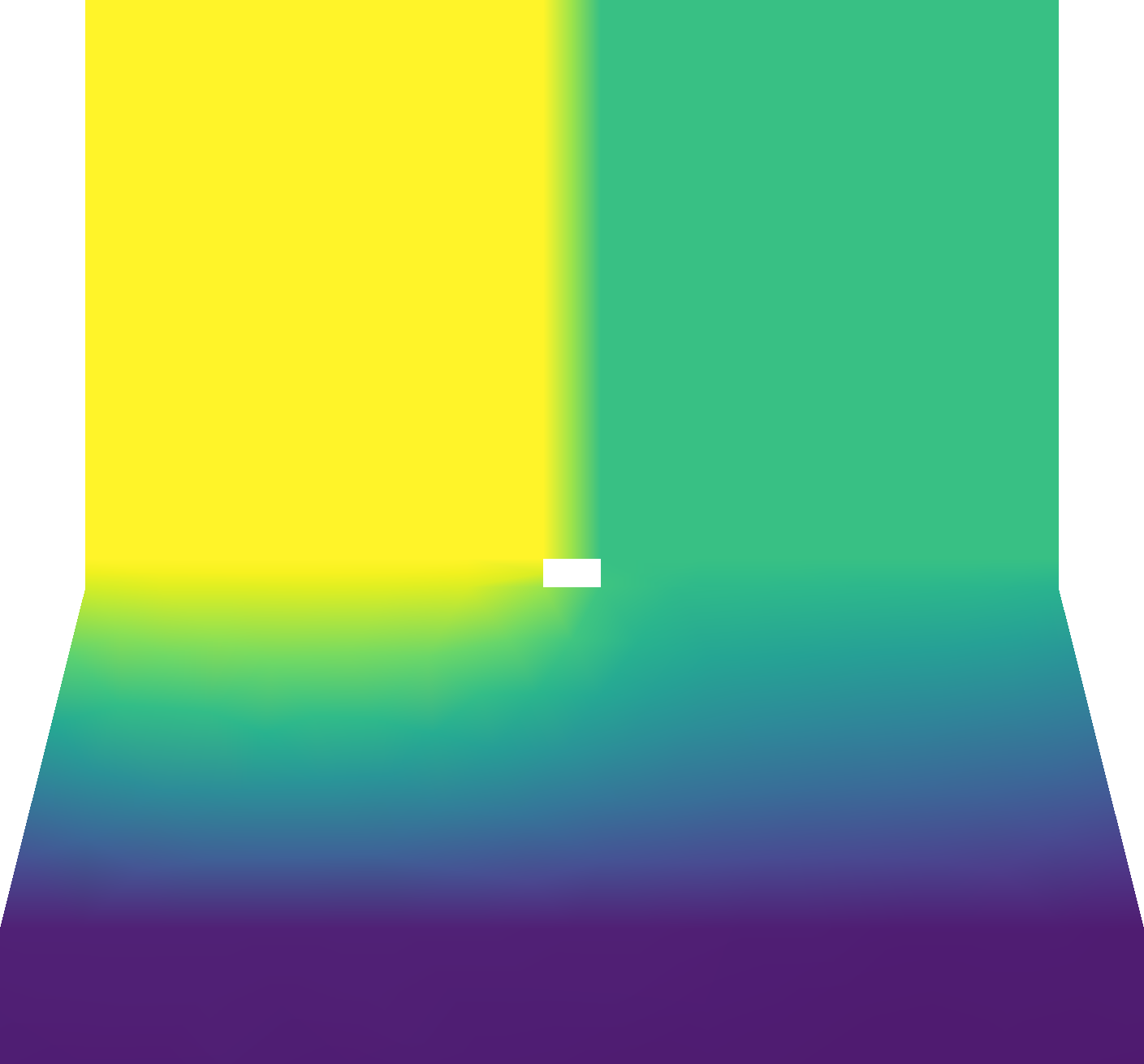}%
		\caption{Mortar TSA solution.}%
	\end{subfigure}
	\caption{Comparison of reference and mortar TSA solution at time $t = \SI{2}{\second}$.}
	\label{hahn:fig_solution}
\end{figure}

As a reference, we use a conforming FE model with meshed insulation and a mesh size of $\SI{0.1}{\milli\meter}$ in the entire domain. The corresponding solution is shown in Fig.~\ref{hahn:fig_solution}.
To illustrate non-conforming meshes on both sides of the interface, the mortar TSA model uses a mesh size of $\SI{0.1}{\milli\meter}$ in the right cable and $\SI{0.25}{\milli\meter}$ in the rest of the domain.
The TSA for the insulation layers indicated in Fig.~\ref{hahn:fig_example} consists of $N=3$ shells, which use the finer interface mesh to obtain a single Lagrange multiplier space.
The cables are almost isothermal due to their high thermal conductivity. Temperature gradients appear mostly across insulation layers. Figure~\ref{hahn:fig_tmax} shows the maximum temperature $T_\mathrm{max}$ in the right cable of the reference solution over time. A stationary constant temperature is reached due to a balance between constant heating $Q$ and the cryogenic cooling condition.
The relative error of the mortar TSA solution compared to the reference solution is shown in Fig.~\ref{hahn:fig_error} with excellent agreement between the two models.
\begin{figure}[t]
	\centering
	\begin{subfigure}[t]{0.49\textwidth}
		\centering
		\includegraphics[width=\textwidth]{tmax}%
		\caption{Maximum temperature.}%
		\label{hahn:fig_tmax}
	\end{subfigure}
	\hfill
	\begin{subfigure}[t]{0.49\textwidth}
		\centering
		\includegraphics[width=\textwidth]{error}%
		\caption{Relative error of mortar TSA solution.}%
		\label{hahn:fig_error}
	\end{subfigure}%
	\caption{Maximum temperature $T_\mathrm{max}$ in the right cable. Reference solution and relative error over time.}%
	\label{hahn:fig_tmax_err}
\end{figure}
Both mortar TSA and reference model are implemented in the free and open-source FE framework \textsc{GetDP}~\cite{GeuzaineGetDP}, using material property functions provided by~\cite{cern-material-library}. The implementation of the problem is publicly available at~\cite{repo}.
\section{Conclusion}
In this work, the concepts of mortar methods and thin shell approximations have been combined to present the mortar TSA technique. It alleviates the need for conforming meshes in the TSA. The problem formulation has been derived for the heat equation and implemented in the free and open-source finite element framework \textsc{GetDP}, with the source code publicly accessible at~\cite{repo}. Good agreement between the method and a classical conforming finite element solution with a meshed insulation layer was shown for a two dimensional non-linear model problem of a simplified superconducting accelerator magnet.
Future research may address a mathematical analysis of the formulation,
study the effect on T- and X-shape thin shell geometries and the extension to magneto-quasistatic formulations.

\paragraph{Acknowledgement}
	The work of R. Hahn and E. Schnaubelt is supported by the Graduate School CE within the Centre for Computational Engineering at TU Darmstadt. The work of E.Schnaubelt is further supported by the Wolfgang Gentner Programme of the German Federal Ministry of Education and Research (grant no. 13E18CHA).

\begin{figure}[t]
	\centering
	\begin{minipage}[t]{0.42\textwidth}
		\strut\vspace*{-\baselineskip}\newline
		\includegraphics[width=1.\columnwidth]{debug_example}
	\end{minipage}
	\hfill
	\begin{minipage}[t]{0.55\textwidth}
		\caption{Simplified accelerator magnet model with temperature-dependent material properties (not to scale). Blue lines designate mortar TSA insulation layers. The model heats up over time due to the heat source in the cables.}%
		\label{hahn:fig_example}
	\end{minipage}
	\vspace{-0.5cm}
\end{figure}

%
%
\bibliographystyle{spmpsci}

%
\begin{figure}[h]
	\vspace{-0.5cm}
	\centering
	\includegraphics[width=0.9\textwidth]{control_line}%
	\caption{Solution along a horizontal control line in the middle of the cables at vertical position $y=\SI{5.5}{\milli\meter}$ and time $t=\SI{2}{\second}$ crossing the insulation layer. Both solutions are in good agreement overall, with a maximum relative error of $\SI{7.31e-4}{}$. Dashed lines indicate the boundaries between cables and insulation.}
	\label{hahn:fig_control_line}
\end{figure}
\end{document}